%% file: PMM-CommsBio.tex
\documentclass[reprint,aip,jcp]{revtex4-1}
\draft

\input{header.tex}

\newcommand{\red}[1]{{\color[rgb]{0,0,0} {#1}}}

\begin{document}
	
\title{Stochastic modelling reveals mechanisms of metabolic heterogeneity} 



\author{Mona K. Tonn}
\affiliation{Department of Mathematics, Imperial College London, SW7 2AZ, London, United Kingdom}

\author{Philipp Thomas}
\affiliation{Department of Mathematics, Imperial College London, SW7 2AZ, London, United Kingdom}

\author{Mauricio Barahona}
\affiliation{Department of Mathematics, Imperial College London, SW7 2AZ, London, United Kingdom}

\author{Diego A. Oyarz\'un}
\email[]{Corresponding author: d.oyarzun@ed.ac.uk}
\affiliation{School of Informatics, University of Edinburgh, EH8 9AB, Edinburgh, EH8 9AB, United Kingdom}
\affiliation{School of Biological Sciences, University of Edinburgh, EH9 3JH, Edinburgh, United Kingdom}


\begin{abstract}
Phenotypic variation is a hallmark of cellular physiology. Metabolic heterogeneity, in particular, underpins single-cell phenomena such as microbial drug tolerance and growth variability. Much research has focussed on transcriptomic and proteomic heterogeneity, yet it remains unclear if such variation permeates to the metabolic state of a cell. Here we propose a stochastic model to show that complex forms of metabolic heterogeneity emerge from fluctuations in enzyme expression and catalysis. The analysis predicts clonal populations to split into two or more metabolically distinct subpopulations. We reveal mechanisms not seen in deterministic models, in which enzymes with unimodal expression distributions lead to metabolites with a bimodal or multimodal distribution across the population. Based on published data, the results suggest that metabolite heterogeneity may be more pervasive than previously thought. Our work casts light on links between gene expression and metabolism, and provides a theory to probe the sources of metabolite heterogeneity.
\end{abstract}


\maketitle

\section*{Introduction}

Cellular heterogeneity is ubiquitous across all domains of life. In microbes, clonal populations display phenotypic variability as a result of multiple factors such as fluctuations in the microenvironment, stochasticity in gene expression, or asymmetric partitioning at cell division \cite{Elowitz2002, Paulsson2005, Raj2008}. Variability is well recognised at the transcriptional and translational levels. Yet various single-cell phenomena result from the emergence of distinct metabolic states within a clonal population. For example, metabolic heterogeneity plays a key role in antibiotic tolerance\cite{Balaban2004, Lewis2007, Shan2017}, heterogeneous nutrient uptake \cite{Vilhena2018, Nikolic2017}, and variations in growth rate \cite{Kiviet2014,thomas2018sources}. It has also been shown that nutrient shifts can cause populations to split into two \cite{vanHeerden2014, Kotte2014} or more \cite{Simsek2018} subpopulations with distinct growth abilities. The emergence of subpopulations has been theorised as a bet-hedging strategy that gives an evolutionary advantage for survival in adverse environments \cite{Balaban2004,Acar2005}. 

A central challenge to quantify metabolic variability is the lack of techniques for measuring metabolites with single-cell resolution\cite{Takhaveev2018}. In contrast to single-cell measurements of protein expression, for which powerful reporter systems have been developed \cite{Golding2005, Taniguchi2010}, quantification of metabolites in single-cells remains a major challenge. Some of the techniques employed so far include F\"orster resonance energy transfer (FRET) sensors \cite{Lemke2011}, metabolite-responsive transcription factors \cite{Xiao2016, Mannan2017}, RNA sensors \cite{Paige2012}, and mass-spectrometry \cite{Ibanez2013}, yet most of these technologies are in early stages of development\cite{Takhaveev2018}. As a result, metabolic heterogeneity is typically quantified indirectly via measurements of metabolic enzymes or growth rate in single-cells \cite{Ozbudak2004,Kotte2014,Kiviet2014}. 

Our objective in this paper is to characterise heterogeneity in metabolites as a result of stochastic enzyme expression and catalysis. Metabolic models traditionally assume that enzymatic reactions behave deterministically on the basis that both enzymes and metabolites appear in high molecule numbers \cite{cornish-bowden04a}. However, single-cell proteomics in \emph{Escherichia coli} show that metabolic enzymes are as variable as any other member of the proteome \cite{Taniguchi2010}, while metabolomics data suggest that average metabolite abundances span several orders of magnitude\cite{Bennett2009}. The few datasets on single-cell metabolite abundance already suggest substantial variability in some metabolites in \emph{E. coli} \cite{Yaginuma2014, Xiao2016}. Such evidence casts doubt on the traditional assumption of metabolism being a purely deterministic process, suggesting a link between fluctuations in enzyme expression and metabolites.

The role of stochastic gene expression in protein variability has been well studied~\cite{Paulsson2005, Raj2008, Shahrezaei2008, Labhsetwar2013, Thomas2014}, but the impact of such randomness on metabolic reactions remains much less understood. Various theoretical studies have analysed the impact of fluctuations in the supply and consumption of metabolites \cite{Levine2007, Thomas2011, Gupta2017}, or the propagation of enzyme noise to a metabolite \cite{Oyarzun2015}. However, despite the advancing experimental evidence of stochasticity in metabolism, mathematical models still lack the sufficient detail to integrate the processes that are known to shape protein heterogeneity, such as stochastic promoter switching and transcriptional bursting.

In this paper we propose a model for metabolite heterogeneity in single-cells. The model integrates stochasticity in enzyme catalysis \cite{cornish-bowden04a} and expression \cite{Shahrezaei2008}, two well-established processes that so far \red{have been studied in isolation}. Our approach includes a stochastic formulation of various relevant mechanisms in enzymatic reactions, including reversible catalysis, stochastic switching of promoter activity, fluctuations in mRNA transcripts, and consumption of the enzymatic product by downstream processes. 

We probe the model for various sources of stochasticity using simulations and analytical solutions for the stationary distribution of the metabolite. The analysis reveals intricate patterns of heterogeneity that translate into bimodal and multimodal distributions for the number of metabolite molecules. These phenomena arise from the interplay between a lowly abundant enzyme and its catalytic parameters. Under the separation of timescales typical of metabolic reactions, we show that metabolite distributions can be accurately approximated by a Poisson Mixture Model across large regions of the parameter space. The mixture model can be readily adapted to a wide class of gene expression models and provides a quantitative tool to predict metabolite variability from enzyme measurements in single-cells.

\section*{Results}

\begin{figure*}
\includegraphics[scale = 0.95]{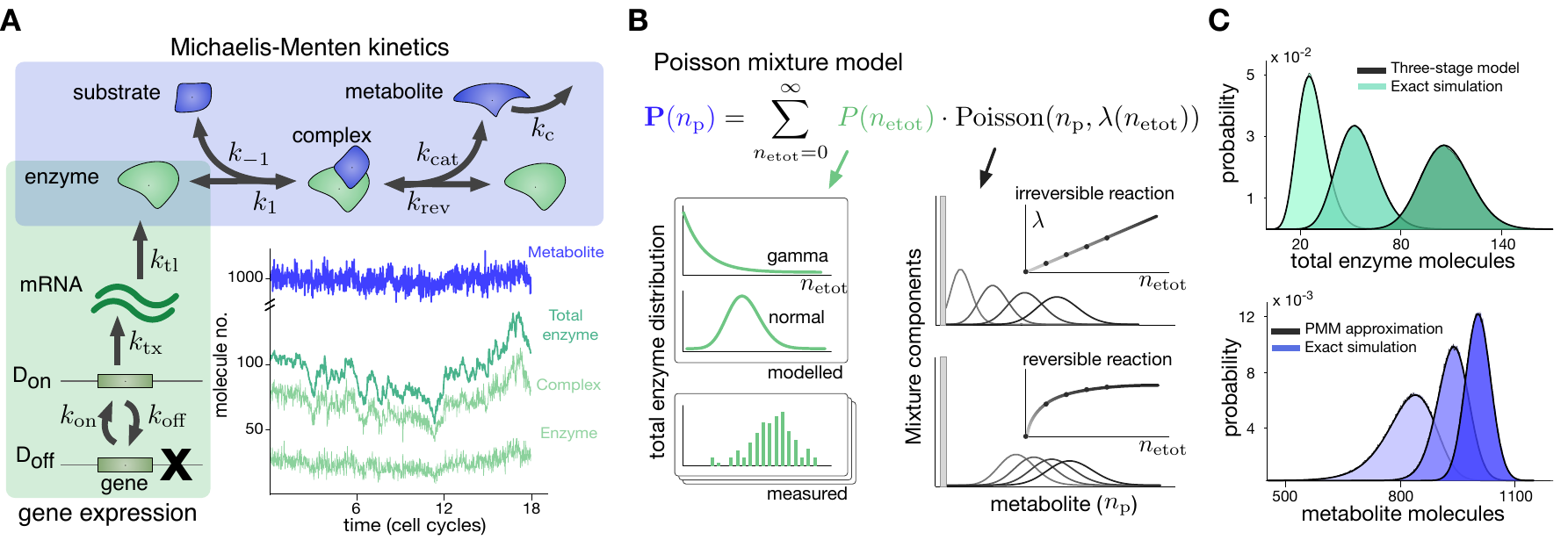}
\caption{\textbf{Stochastic model for an enzymatic reaction.} \textbf{(A)} The model integrates reversible Michaelis-Menten kinetics with the three-stage model for gene expression \cite{cornish-bowden04a,Shahrezaei2008}. The model includes consumption of the metabolite by downstream pathways, degradation of mRNA transcripts, and dilution of all chemical species by cell growth (not shown in diagram); rate constants are shown in the figure and model reactions are shown in equations \eqref{substrate_binding}--\eqref{molecule_dilution2} of the Methods. The inset shows a typical simulation for a realistic parameter set shown in Table \ref{parameters_model1}. \textbf{(B)} Construction of the Poisson Mixture Model (PMM) for the number of metabolite molecules ($\product$). This approximation is valid under a separation of timescales between enzyme expression and enzyme catalysis. The mixture model, shown in equation \eqref{PoissonMixtureModel}, comprises Poisson distributions weighted by the distribution of enzyme expression $\mathbf{P}(\etotal)$. The Poisson parameter $\lambda(\etotal)$ depends on enzyme kinetics via the nonlinear relation in equation \eqref{lambda}. In the irreversible case ($\krev=0$), the $\lambda(\etotal)$ parameter scales linearly and produces equi-spaced Poisson modes. The mode $\text{Poisson}(\product,0)$ is highlighted as a bar. \textbf{(C)} The PMM provides an accurate approximation of the stationary distributions. Insets show distributions for enzyme and metabolite, computed via Gillespie simulations and the PMM approximation for fixed $\lambda_{\infty}=\text{1080}\,\text{molecules}$, $K=\text{8}\,\text{molecules}$, and three different promoter switching parameters, shown in Table \ref{parameters_model1}.}
\label{model_intro}
\end{figure*}

\subsection*{Stochastic model of an enzymatic reaction}

We consider a model that combines enzyme kinetics and enzyme expression into a single stochastic description (Figure \ref{model_intro}A). The model includes an enzymatic reaction with standard Michaelis-Menten kinetics, in which substrate and enzyme bind reversibly to form a complex that undergoes reversible catalysis into a metabolite. We assume that enzyme expression follows the well established three-stage model for gene expression~\cite{Raj2008,Shahrezaei2008}, where a single copy gene switches stochastically between an inactive state ($\doff$) and active state ($\don$). In the active state, mRNAs are transcribed and translated into protein. The model also includes consumption of the metabolite by downstream pathways, degradation of mRNA transcripts, and dilution by growth of all species. Since metabolic reactions operate far from thermodynamic equilibrium, we assume that the substrate pool remains constant so that the system reaches a non-zero flux, \eg when the substrate is a highly abundant extracellular carbon source or a slowly varying intracellular metabolite. The model reactions are shown in equations \eqref{substrate_binding}--\eqref{molecule_dilution2} in the Methods section.

To investigate the emergence of metabolic heterogeneity, we need to compute the stationary probability distribution of metabolite molecules ($\product$) for relevant combinations of model parameters. Figure \ref{model_intro}B shows a typical simulation of the model obtained with Gillespie's algorithm \cite{Gillespie1976}. A key challenge for such simulations, however, is the multiscale nature of enzymatic reactions: not only do metabolic reactions operate in a much faster timescale (milliseconds) than enzyme expression (tens of minutes)\cite{Levine2007, cao2005accelerated, Lugagne2013}, but also the average number of enzymes is much lower than the number of metabolites. These multiple scales result in reaction propensities that differ by several orders of magnitude, thus leading to extremely slow simulations which make the exploration of the parameter space infeasible. An alternative is to use simulation algorithms that exploit the separation of scales to increase computational speed, such as tau-leaping or slow-scale approximations\cite{Gillespie2007}. Yet in our case it is unclear how such numerical approximations impact the predictions drawn from the simulations.

To determine the impact of genetic and catalytic parameters on metabolic heterogeneity, we obtained an analytic approximation for the distribution of metabolite molecules that can be evaluated efficiently without expensive stochastic simulations. Our solution allows the exploration of parameter space to characterise the different regimes promoting metabolic heterogeneity. The approximation follows from exploiting time scale separation in the Chemical Master Equation of the stochastic process~\cite{vanKampen1992}. In physiological regimes the model has three timescales: a fast metabolic time scale, in which substrate and enzyme bind and unbind; an intermediate time scale associated with the catalysis of the metabolite ($\product$); and a slow timescale associated with the expression of the enzyme and dilution by cell growth. 

The total amount of enzyme (free and substrate-bound, denoted as $\enzyme$ and $\complex$ respectively) varies in the slowest timescale, and therefore the binding/unbinding of substrate and enzyme equilibrates quickly. As a result, in the timescale of gene expression, the metabolite can be assumed to depend directly on the the total enzyme $\etotal=\enzyme+\complex$ rather than on $\enzyme$ and $\complex$ individually. Under this approximation, it is convenient to use the law of total probability:
\begin{align}\label{PoissonMixtureModel}
\mathbf{P}(\product) = \sum_{\etotal=0}^{\infty}  \underbrace{\mathbf{P}(\etotal)}_{\text{gene expression}}  \underbrace{\mathbf{P}(\product|\etotal)}_{\text{catalysis}}.
\end{align}

The formula in \eqref{PoissonMixtureModel} decomposes the distribution of metabolite $\mathbf{P}(\product)$ into stochasticity originating from enzyme expression, $\mathbf{P}(\etotal)$, and from fluctuations in the catalytic reaction itself, described by the conditional distribution of metabolite given the amount of total enzyme, $\mathbf{P}(\product|\etotal)$. In the timescale of metabolite fluctuations, the total enzyme can be assumed to be in a quasi-stationary state. Further, exploiting the fast binding/unbinding between substrate and enzyme, we showed that the metabolite follows a birth-death process with effective propensities (details in Methods section):
\begin{align}
\begin{split}
	\text{k}_{\text{birth}}^{\text{eff}} &= \mathrm{\kcat} \mathbb{E}(\complex  |\etotal,\product)\\
	& \approx  \mathrm{\kcat} \frac{ \kback }{(\kback+\kfor \substrate)} \etotal,\\
	\text{k}_{\text{death}}^{\text{eff}} &= \krev \mathbb{E}(\enzyme  |\etotal,\product) + \kcons\\
	& \approx \krev \frac{ \kfor \substrate }{(\kback+\kfor \substrate)} \etotal + \kcons,
	\end{split}\label{eff_propensities_main}
\end{align}
where $\mathbb{E}(\enzyme  |\etotal,\product)$ and $\mathbb{E}(\complex  |\etotal,\product)$ are the conditional expectations of the free enzyme ($\enzyme$) and complex ($\complex$) given the total enzyme and metabolite. In equation \eqref{eff_propensities_main}, $n_{s}$ is the constant number of substrate molecules, the parameters $\kfor$, $\kback$, $\kcat$, and $\krev$ are the rate constants of the Michaelis-Menten mechanism (defined in \figref{model_intro}A), and $\kcons$ is an effective first-order rate constant of metabolite consumption by downstream pathways. The conditional distribution needed in the model \eqref{PoissonMixtureModel} can then be computed explicitly:

\begin{align}\label{Metabolite_conditional_distribution}
\mathbf{P}(\product|\etotal) & \sim \text{Poisson} \left(\product; \lambda(\etotal)\right),
\end{align}
with Poisson parameter  
\begin{align}
\label{lambda}
\lambda(\etotal) &= \frac{\lambda_{\infty}}{1+K\slash\etotal},
\end{align}
and $(\lambda_{\infty},\,K)$ are two effective kinetic parameters
\begin{align}
\lambda_{\infty} = \substrate\frac{\kcat \kfor}{\krev \kback},\,\text{and} \,\,K &= \kcons\frac{\kfor \substrate + \kback}{\krev \kback}.\label{effparameters}
\end{align}
The parameters $\lambda_{\infty}$ and $K$ are in units of molecules$\slash$cell and depend on the interplay between substrate abundance, enzyme kinetics, and downstream processes.

As illustrated in \figref{model_intro}B, the distribution in \eqref{PoissonMixtureModel} is a Poisson Mixture Model\cite{Chaturvedi1977,Iyer-Biswas2009,Dattani2016} (PMM) that convolves the enzyme distribution $\mathbf{P}(\etotal)$ with various Poisson modes $\mathbf{P}(\product|\etotal)$ arising from the catalytic activity. In our model, the analytical distribution of the total enzyme abundance follows the standard solution of the three-stage model for gene expression \cite{Shahrezaei2008}, which can be computed explicitly in terms of model parameters. In certain limits, the three-stage model produces approximately Gamma or Normal distributions depending on the the mean expression level and the half lives of mRNAs and proteins\cite{Raj2008,Shahrezaei2008}.

The decomposition in equation \eqref{PoissonMixtureModel} shows that the PMM is not limited to the model for gene expression we have considered here. Other models may be used, either by using closed-form expressions for $\mathbf{P}(\etotal)$, or by inferring the enzyme distribution directly from single-cell protein expression data such as flow cytometry or single-cell microscopy \cite{Taniguchi2010, Elowitz2002}. The PMM thus provides a versatile tool to predict metabolite heterogeneity from modelled or measured enzyme heterogeneity viewed as an upstream source of variation\cite{Dattani2016}. \figref{model_intro}C shows that the PMM distribution provides a good approximation to Gillespie simulations computed with typical parameter values.

\subsection*{Qualitative features of the Poisson Mixture Model}

\begin{figure*}
\centering
\includegraphics[scale=1]{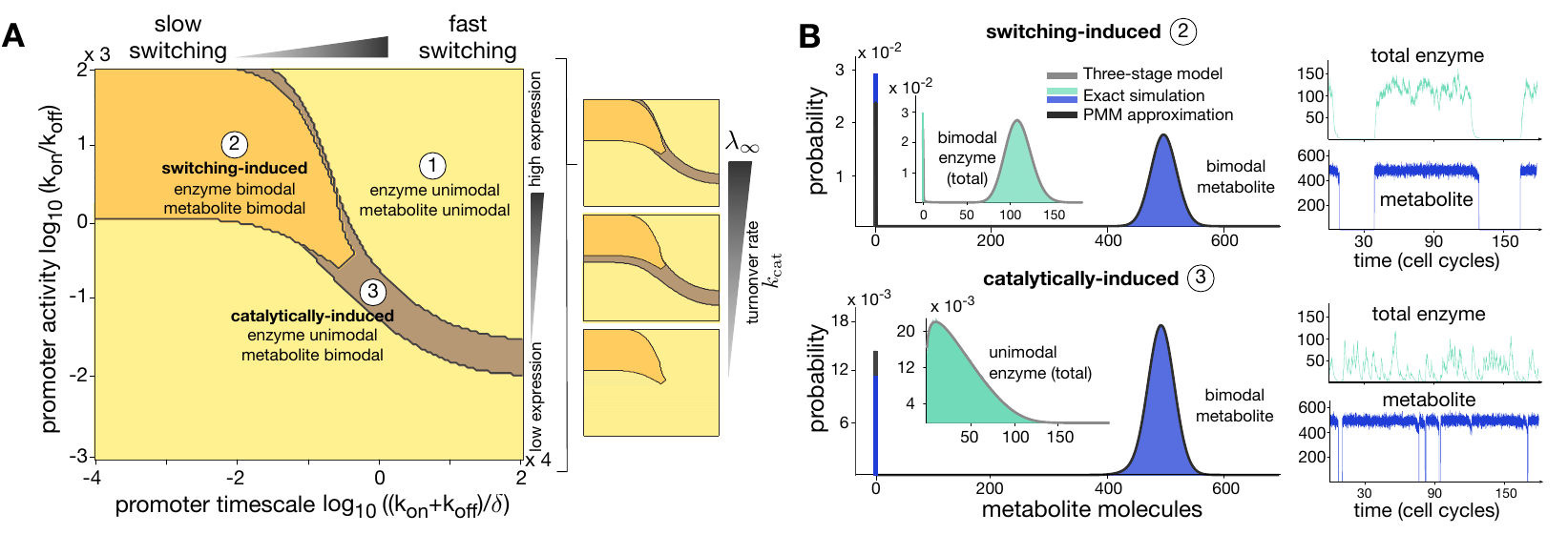}
\caption{\textbf{Mechanisms for metabolite bimodality.}
 \textbf{(A)} We evaluated the Poisson Mixture Model across a broad range of promoter switching timescale 
and promoter activity. 
Unimodal distributions for enzyme and metabolite (similar to those shown in Figure \ref{model_intro}C) cover a large fraction of the parameter space. We identified two regimes in which metabolites are bimodal: in the switching-induced regime, bimodality propagate from the enzyme to metabolite. In the catalytically-induced regime, bimodality originates from a lowly abundant enzyme and the strong separation of timescales between expression and catalysis. The small panels show model predictions for a fixed kinetic parameter $K = 0.1333$ molecules, and increasing $\lambda_{\infty} = \left\{ 300,\,3000,\,30000 \right\}\,\text{molecules}$, obtained by increasing the turnover rate constant $\kcat$. \textbf{(B)} Exact simulations for two parameter sets verify the predictions drawn from the PMM approximation. We simulated over a long time horizon to obtain accurate estimates for stationary distributions; insets shown only a small portion of the time courses. The parameter values for the promoter switching rates are indicated in panel A and we fixed $\lambda_{\infty}=500$ molecules. Both types of bimodality can be clearly distinguished in the time courses, but we note that they lead to almost identical distributions for the metabolite. In both cases, the PMM provides an accurate approximation for the stationary distributions.}
\label{bimodality_enzyme_product}
\end{figure*}

At the heart of the metabolite PMM is the interplay between variability from gene expression and that originating from enzyme kinetics. Specifically, the Poisson parameter $\lambda(\etotal)$ in equation \eqref{lambda} controls the density and dispersion of the Poisson modes, which in turn shape the overall pattern of variability. As shown in \figref{model_intro}B, there are several cases of interest. For example, for irreversible reactions ($\krev=0$), the Poisson parameter simplifies to 
\begin{align}
\lambda(\etotal) = \frac{\kfor \substrate \kcat}{(\kfor \substrate + \kback)\kcons}\etotal,
\end{align}
which scales linearly with the enzyme abundance and thus the Poisson modes have equidistant means. In reversible reactions, on the other hand, the Poisson parameter saturates and causes the Poisson modes to concentrate around  $\lambda_{\infty}$. This effect is stronger for strong reversibility (high $\krev$), in which case the kinetic parameter $K$ is small. Note also that in either case, as the enzyme number $\etotal$ grows, the Poisson modes spread out since $\lambda(\etotal)$ controls both their mean and variance.

From the construction of the PMM in \eqref{PoissonMixtureModel}, we observe that the enzyme distribution weighs the various Poisson modes, potentially producing metabolite distributions that are unimodal, bimodal or even multimodal. For example, for highly expressed reversible enzymes, the distribution $\mathbf{P}(\etotal)$ is non-negligible when $\etotal$ is large. Hence most Poisson modes do not contribute to the final metabolite distribution, except the mode centred at $\lambda_{\infty}$, which leads to a unimodal metabolite distribution with a mean close to the deterministic average. 

Conversely, for lowly expressed enzymes, there is a non-negligible probability of enzymes not being expressed, and thus the first term of the PMM, \ie $\mathbf{P}(0)\text{Poisson}(\product,0)$, causes the metabolite distribution to peak at zero. However, the metabolite distribution may also display a second peak at $\lambda_{\infty}$ if, for example, the $\lambda(\etotal)$ parameter causes many Poisson modes to concentrate around $\lambda_{\infty}$. This results in a bimodal metabolite distribution, whereby an isogenic population splits into metabolite producers and non-producers. Similar reasoning can be used to understand the emergence of multimodal metabolite distributions, which correspond to three or more subpopulations with varying metabolic activities. This qualitative analysis suggests that metabolic subpopulations can emerge even in cases where enzymes display unimodal distributions across the population. Crucially, this also indicates that metabolic subpopulations emerge through mechanisms that do not follow trivially from transcriptional heterogeneity alone, as we explore in more detail in next section.

\subsection*{Mechanisms for metabolic bimodality}

\begin{figure*}
\includegraphics[scale=1]{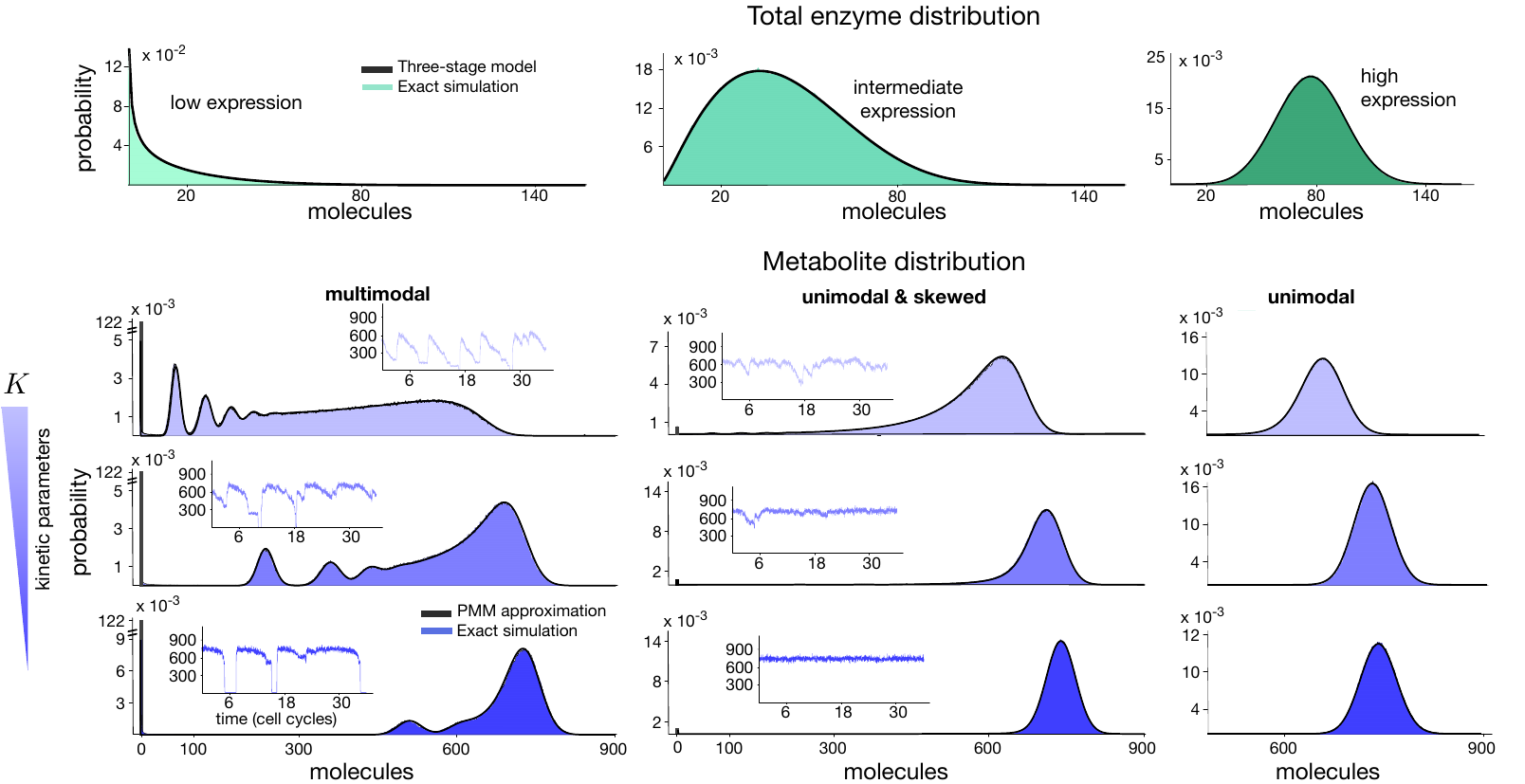}
\caption{\textbf{Emergence of metabolic multimodality.} 
 We used the PMM approximation to find regimes for multimodality through perturbations to the enzyme kinetics. We vary the kinetic parameter $K$ to control the dependency of the Poisson parameter $\lambda(\etotal)$ in equation \eqref{lambda} on the total enzyme abundance. Parameter values are $\lambda_{\infty} = 750\,\text{molecules}$ and $K = \left\{ 10.0400,\,2.1630,\,    0.4660 \right\}\,\text{molecules}$ obtained by variations to the kinetic rate constants $\kcat$ and $\krev$ with a constant ratio $\kcat\slash\krev$.
We shape the mean enzyme abundance with the
promoter switching rates $\kon = \left\{1.56, 5.9, 20\right\}\times10^{-4}\text{ s}^{-1}$ and $\koff = \left\{9.8, 9.3, 8\right\}\times10^{-4}\text{ s}^{-1}$. From the PMM we found intricate patterns of multimodal distributions in the metabolite, all of which show an excellent match with the corresponding Gillespie simulations. The simulated time courses show metabolite numbers traversing various quasi-stationary regimes.
}
\label{metabolite_multimodal}
\end{figure*}

First, we explored the impact of stochastic promoter switching on the emergence of metabolite bimodality. \figref{bimodality_enzyme_product}A shows the summary of calculations when evaluating the PMM for variations in two key characteristics of the promoter across several orders of magnitude: the switching time scale ($(\kon+\koff)\slash\delta$) and the promoter activity ($\kon\slash\koff$), for various values of the kinetic parameter $\lambda_{\infty}$. We found three qualitatively distinct parameter regimes for the metabolite distribution that emerge from the combination of stochastic switching and catalysis: (1) a regime where both enzyme and metabolite have unimodal distributions, akin to the results shown earlier in \figref{model_intro}C; (2) a regime where both enzyme and metabolite have bimodal distributions; and (3) a regime in which the enzyme is unimodal but the metabolite is bimodal.

It can be shown that the deterministic version of our model in equations \eqref{substrate_binding}--\eqref{molecule_dilution2} has a single steady state. Hence
regime (1) can be thought of as a stochastic correction consisting of unimodal distributions around a deterministic steady state. This is the expected behaviour under the traditional assumptions of high abundance of enzyme and metabolite molecules. 

The other two regimes, however, correspond to alternative routes of noise-induced bimodality that cannot be explained using deterministic models\cite{Oyarzun2015b, Lipshtat2006, To2010}. Regime (2) is a highly stochastic regime dominated by the slow stochastic switching of the promoter, which drives and entrains the metabolic response. Hence we term it \emph{switching-induced bimodality}. Slowly switching promoters are known to produce bimodal gene expression~\cite{Thomas2014, Dattani2016}, and thus this regime corresponds to a case in which bimodality propagates from enzymes to metabolites. \figref{bimodality_enzyme_product}A shows that this behaviour appears robustly for slow switching and high promoter activity across values of $\lambda_{\infty}$ kinetic parameter.

Regime (3), the second route for metabolite bimodality, originates from a unimodal but weakly expressed enzyme (low $\kon\slash\koff$) expressed from fast switching promoters. In this case, the birth of a small number of enzyme molecules is sufficient to kick-start catalysis and make it rapidly settle in a quasi stationary regime. This distinct phenomenon is a result of the separation of time scales between enzyme expression and catalysis, and we refer to it as \emph{catalytically-induced bimodality}. From Figure \ref{bimodality_enzyme_product}A, we observe that this form of bimodality appears for a narrow range of promoter switching parameters corresponding to fast switching genes with medium to low promoter activity. This behaviour disappears altogether for a low $\lambda_{\infty}$ parameter, for example in case of strong reversibility. 

To validate the predictions of the PMM approximation, we ran full Gillespie simulations over a long time horizon for different parameter sets. Figure \ref{bimodality_enzyme_product}B shows the simulation time courses and resulting histograms. For switching-induced bimodality, we observe how slowly switching promoters causes a single cells to lack the enzyme over several cell cycles, a period during which the metabolite is not produced. In the case of catalytic-induced bimodality, however, fast switching combined with a low average expression level cause the metabolite abundance to drop for shorter but more frequent intervals. In both cases the PMM provides an excellent approximation to the bimodal histograms obtained from the stochastic simulations. Furthermore, we observe that the bimodal metabolite distributions both regimes are almost indistinguishable from each other, yet they are produced by enzymes with substantially different time courses and distributions. These regimes therefore correspond to distinct forms of bimodality, arising from fundamentally different mechanisms. 

\subsection*{Emergence of metabolic multimodality}

To explore the emergence of multimodality, we examined the analytical formula of the PMM in \eqref{PoissonMixtureModel} to identify kinetic regimes associated with distinct enzyme distributions. A necessary condition for the emergence of multiple modes is that the Poisson components do not overlap and are sufficiently spaced from each other. From the definition of the $\lambda(\etotal)$ parameter in \eqref{effparameters}, this happens when the kinetic parameter $K$ is large. As discussed earlier, depending on the distribution of the enzyme, the Poisson modes may appear or cancel in the final metabolite distribution. We thus swept the parameter $K$ and evaluated the PMM across various enzyme expression levels, including low expression with a skewed distribution and high expression with a normally distributed enzyme.

As shown in Figure \ref{metabolite_multimodal}, we found intricate patterns of multimodal distributions, depending on the interplay between the heterogeneity of the enzyme, $\mathbf{P}(\etotal)$, and the enzyme kinetics encapsulated by the $K$ parameter. Multimodality appears when the enzyme expression levels are low as compared to the parameter $K$. For instance, the values of $K$ in \figref{metabolite_multimodal} are approximately 5-, 20-, and 100-fold those used in the bimodal examples in \figref{bimodality_enzyme_product}. 
For enzymes expressed at intermediate levels, in the order of tens of molecules/cell on average, we found metabolite distributions that are unimodal but highly skewed. In the case of highly expressed enzymes, metabolites followed approximately normal distributions for a wide range of kinetic parameters.  

The predictions are confirmed by Gillespie simulations of the full stochastic model, which display a striking match with the PMM approximation, even for complex multimodal distributions. 
The simulation time courses (shown in the insets of \figref{metabolite_multimodal}) show that the multiple modes for weakly expressed enzymes correspond to cells remaining in a fixed metabolic state over the scale of the cell cycle but fluctuate across other states over longer time scales. For intermediate enzyme expression and large values for $K$, the metabolite does not settle in the quasi-stationary states and displays a long-tailed distribution. A decrease in $K$ suppresses the tail of the distribution driving the PMM towards an approximately normal distribution. Altogether, these results indicate that the relation between enzyme expression and the kinetic parameters $\lambda_{\infty}$ and, in particular, $K$ are key determinants for the emergence of multimodality. This underscores the utility of the PMM to guide the prediction of qualitative and quantitative features of metabolite distributions for a wide range of parameter combinations.

\section*{Discussion}
Metabolic reactions are the powerhouse of living systems, fuelling the activity and dynamics of most cellular functions. Yet metabolism has been traditionally considered as a static process isolated from the rest of the cellular machinery. 
\red{Currently, the accepted notion is that due to the large number of molecules involved, metabolism is a deterministic process at the cellular level, modulated by potentially random extrinsic factors\cite{Wehrens2018,Kotte2014}.} Here we integrated enzyme kinetics and enzyme expression to propose a theoretical model for variability of metabolites in single cells. \red{The model suggests that cell-to-cell metabolite variation can also arise as a result of intrinsic sources such as stochastic fluctuations in enzyme expression. }


The majority of work on non-genetic heterogeneity has focused on stochastic gene expression and the resulting variability in protein levels \cite{Elowitz2002, Paulsson2005}. This has produced a wealth of single-cell data and models to understand the variability in transcription and translation observed in clonal populations. Metabolite heterogeneity, however, remains poorly understood theoretically and has been observed only indirectly (e.g., through measurements of metabolic enzymes \cite{Ozbudak2004, Kotte2014} or growth rate \cite{Kiviet2014}) due to the lack of techniques to measure metabolite abundance in single cells. 

Using the separation of time scales characteristic of metabolic reactions, we found that the stationary distribution of a metabolite follows a Poisson Mixture Model (PMM). The PMM can be efficiently evaluated across large domains of the parameter space and provides excellent approximations to the distributions computed from full stochastic simulations. Importantly, the model can be readily adapted to include different stochastic models for enzyme expression, beyond the three-stage model considered here~\cite{Raj2008, Corrigan2016}, or even stochastic and time-dependent enzyme expression modelled as upstream drives\cite{Dattani2016}. The model can also be parameterised from experimentally measured distributions for enzyme levels in single-cells\cite{Taniguchi2010}. In combination with the enzyme kinetic parameters, the PMM could provide a powerful tool to predict metabolite variability from single-cell protein data obtained with established techniques such as flow cytometry or time-lapse microscopy.

We found complex patterns of metabolite heterogeneity depending on the interplay between the timescale of promoter activation/deactivation, the enzyme expression level, and the enzyme kinetics. The model predicts that bimodal and multimodal metabolite distributions can emerge in various parameter regimes. In such regimes single-cells spend several cell cycles in a constant metabolic state, but in timescales as long as tens of cell cycles, they switch stochastically across different states. Such long-term fluctuations in single cells result in highly heterogeneous populations containing several subgroups of metabolically distinct cells.

Bimodal metabolic phenotypes have been observed as a result of transcriptional regulation \cite{Ozbudak2004, Acar2005}, post-translational control \cite{vanHeerden2014} and stochastic effects triggered by environmental shifts \cite{Kotte2014}. Our model reveals two distinct regimes in which metabolites display bimodality. One regime, which we call switching-induced bimodality, corresponds to the intuitive case in which a bimodal enzyme produces a bimodal metabolite. In agreement with previous studies on stochastic gene expression, this type of bimodality appears as a result of slow switching between promoter states\cite{Thomas2014,Ge2018}. In addition, we identified a fundamentally different mechanism of \emph{catalytically-induced bimodality}, in which a unimodal enzyme produces a bimodal distribution of metabolite. This phenomenon results from a combination of slow fluctuations of a weakly expressed enzyme and the comparatively faster timescale of enzyme catalysis. Catalytic timescales are typically in the order of seconds or faster, so that slow fluctuations in enzyme expression levels produce two quasi-stationary metabolic states in single cells. At a population level, this leads to two distinct subpopulations of metabolite producers and non-producers. 

\begin{figure}
\centering
\includegraphics{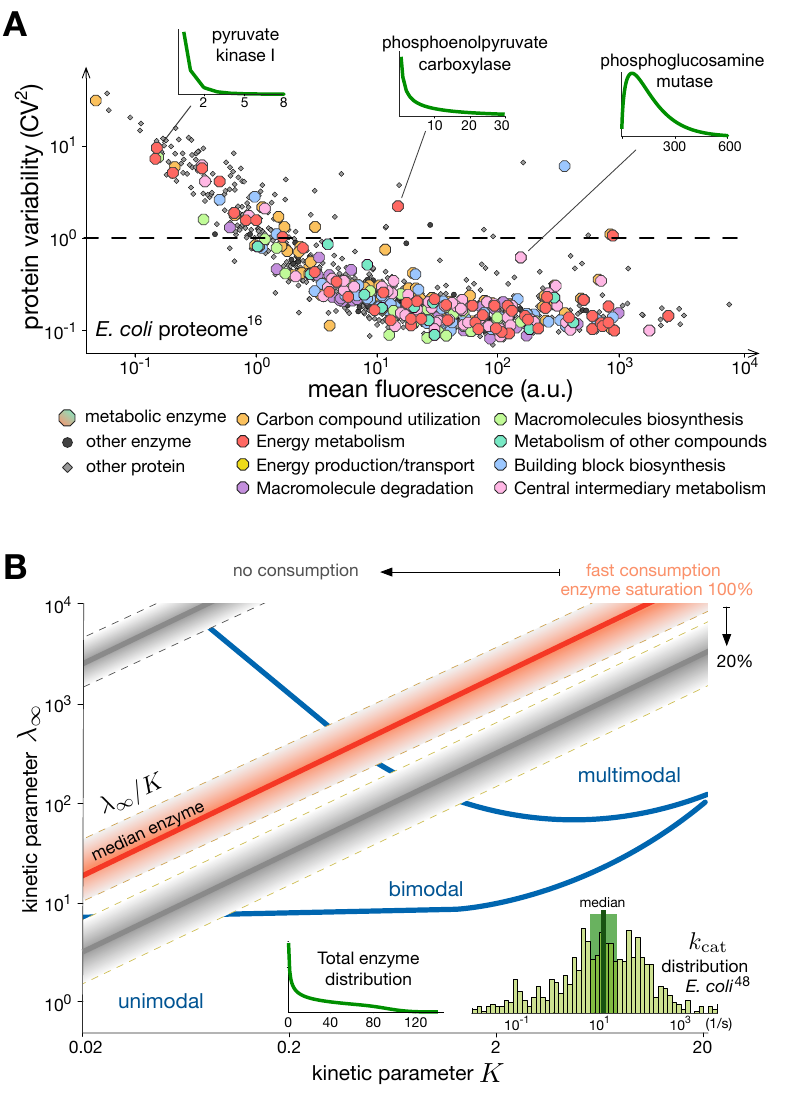}
\caption{\textbf{Model predictions and experimental data.} \textbf{(A)} Single-cell measurements reveal that metabolic enzymes are as variable as other members of the proteome \cite{Taniguchi2010}. Data correspond to $\sim$80\% of the \emph{E. coli} proteome, including 268 enzymes involved in various metabolic functions  \cite{Serres2000} (coloured circles). The coefficient of variation (CV) is defined as standard deviation over mean of measured distributions. Distributions with $\text{CV}>1$ (dashed line) are long-tailed and peak at zero, which resemble the distributions required for catalytically-induced bimodality and multimodality (\figref{bimodality_enzyme_product}B and \ref{metabolite_multimodal}); shown are the distributions of three representative enzymes computed from fitted Gamma distributions \cite{Taniguchi2010}.\textbf{(B)} Predictions of the Poisson Mixture Model for combinations of parameters $\lambda_{\infty}$ and $K$ in \eqref{effparameters}. For a lowly abundant enzyme (distribution in inset), the model predicts unimodality, catalytically-induced bimodality and multimodality in large regions of the $(\lambda_{\infty},\,K)$ space. The red line represents a constant ratio $\lambda_{\infty} \slash K$ for the median $\kcat \approx 16.5 \text{ s}^{-1} $ across 752 enzymes in \emph{E. coli}, see $\kcat$ distribution in inset\cite{Bar-Even2011}. Shaded red corresponds to the lines obtained for $\kcat$ values within a range 0.5- and 2-fold of the median (highlighted in green in the inset). The grey lines and shaded areas correspond to perturbations to the consumption rate constant ($\kcons$) and enzyme saturation ($\epsilon$); more details in Methods.}
\label{figure_parameter_value}
\end{figure}

As shown in \figref{figure_parameter_value}A, single-cell measurements in \emph{E. coli} suggest that metabolic enzymes appear in low copy numbers across most cellular pathways\cite{Taniguchi2010}. In the specific growth conditions of that experiment, the data did not reveal bimodal expression of enzymes, which precludes the emergence of switching-induced bimodality in the metabolites they catalyse. However, as illustrated by the three representative distributions in \figref{figure_parameter_value}A, a number of enzymes have a low mean and a long-tailed distribution, akin to those required for catalytically-induced bimodality and multimodality. This suggests that enzyme distributions found in nature have the characteristics needed for the emergence of subpopulations with two or more distinct metabolite abundances.

Further requirements for metabolite bimodality and multimodality involve conditions on the parameters $\lambda_{\infty}$ and $K$ in equation \eqref{effparameters}. However, their computation requires rate constants ($\kfor$, $\kback$ and $\krev$) that are rarely measured separately, and instead enzymology data typically provides values for $\kcat$ and $\Km = (\kcat+\kback)\slash \kfor$ only \cite{Schomburg2013}. In the Methods we show that the ratio $\lambda_{\infty}\slash K$ can be expressed as $\lambda_{\infty}\slash K = \epsilon \times \kcat\slash\kcons$, where $\epsilon$ is the saturation level of the enzyme and $\kcons$ is the first-order rate constant of metabolite consumption. As illustrated in \figref{figure_parameter_value}B, the ratio $\lambda_{\infty}\slash K$ corresponds to a straight line in a $(\lambda_{\infty},\,K)$-space, and a specific enzyme (\ie with specific values for $\kfor$, $\kback$ and $\krev$) corresponds to a single point on the line. In \figref{figure_parameter_value}B we compare model predictions for a lowly abundant enzyme with different $\lambda_{\infty}\slash K$ ratios computed for $\kcat$ constants measured in \emph{E. coli}\cite{Bar-Even2011}. Considering the large spread in measured $\kcat$ values, of up to seven orders of magnitude, plus the multiple combinations of metabolite consumption rates and enzyme saturation, the analysis suggests that catalytically-induced bimodality and multimodality are plausible within physiological regimes. Further validations of our predictions require measuring metabolite distributions directly, but this is still subject to a number of challenges in single-cell measurement technologies \cite{Takhaveev2018}.

Our analysis shows that metabolite heterogeneity depends on a delicate interplay between enzyme expression and enzyme kinetics. It is reasonable to expect that energy-critical enzymes, such as those in central carbon metabolism, filter away fluctuations through post-translational regulatory mechanisms commonly found in metabolism.  However, this may not be the case in pathways that are dynamically regulated in response to changes in the environment or cellular context. For example, transcriptional regulation in response to nutrient shifts may steer enzyme levels into regimes of low copy numbers where heterogeneity may dominate the resulting phenotypes. Such mechanism has been already shown to produce growth bimodality in the gluconeogenic switch of \emph{E. coli}\cite{Kotte2014}, while a similar mechanism could underpin the large variability observed in single-cell measurements of S-Adenosyl methionine\cite{Paige2012}. Noise-induced phenomena also have implications for the design of dynamic control systems for heterologous pathways, which are focus of much research in synthetic biology and metabolic engineering \cite{Liu2018}.

In our efforts to build a theory that includes components shared by most enzymatic reactions, we have purposely overlooked a number of processes that can shape metabolic activity. For example, we have not addressed the impact of feedback mechanisms that control enzyme activity, including \eg product inhibition and allostery, or transcriptional mechanisms that control enzyme expression in response to metabolites. Since post-translational regulation operates on timescales much shorter than enzyme expression, with similar timescale separation arguments it should be possible to express the metabolite distribution as a mixture model akin to ours. In such case the mixture components are not necessarily Poisson and their distribution will depend on the particular mechanism under study. We expect that bimodal and multimodal responses are likely to emerge in this setting, but the precise parameter conditions would have to be studied on a case-by-case basis. Transcriptional feedback control can also display various mechanisms depending on the particular pathway under study. One common motif relies on transcription factors (TF) that up- or down-regulate enzyme expression upon binding to a specific metabolite\cite{Mannan2017}. These mechanisms have been shown to play important roles on metabolic activity\cite{Chaves2019}, but they also bring to the fore subtle questions that require detailed examination, for example, on the role of fluctuations coming from TF expression itself, or the impact of negative TF autoregulation\cite{Fang2017}. Our study paves the way for these and other questions to be addressed and raises exciting prospects for the future research in metabolic heterogeneity. 

In this paper we laid theoretical foundations to study metabolism in conjunction with stochastic enzyme expression. We brought together classic models for gene expression and enzyme kinetics, and discovered a rich array of distinct stochastic phenomena that underpin the emergence of metabolic subpopulations. Our theory provides a quantitative basis to draw testable hypotheses on the sources of metabolite heterogeneity, which together with the ongoing efforts in single-cell metabolite measurements, will help to re-think metabolism as an active source of phenotypic variation.

\section*{Methods}

\subsection*{Stochastic modelling and simulation}

We built a fully stochastic model for the reaction scheme describing a metabolic reaction coupled with gene expression (Fig.~\ref{model_intro}A):  

\begin{align}
\text{Substrate} + \text{Enzyme} & \mathrel{\mathop{\rightleftarrows}^{\kfor}_{\kback}} \text{Complex} \tag{R1} \label{substrate_binding}\\
\text{Complex} &\mathrel{\mathop{\rightleftarrows}^{\kcat}_{\krev}} \text{Metabolite} + \text{Enzyme} \tag{R2} \label{metabolite_catalysis}\\
\text{mRNA} &\mathrel{\mathop{\rightarrow}^{\ktransl}} \text{mRNA} + \text{Enzyme} \tag{R3} \label{mrna_translation}\\
\text{DNA}_{\text{on}} &\mathrel{\mathop{\rightarrow}^{\ktransc}} \text{DNA}_{\text{on}} + \text{mRNA} \tag{R4} \label{gene_transcription}\\
\text{DNA}_{\text{off}} &\mathrel{\mathop{\rightleftarrows}^{\kon}_{\koff}} \text{DNA}_{\text{on}} \tag{R5} \label{promoter_switching}\\
\text{Metabolite} &\mathrel{\mathop{\rightarrow}^{\kcons}}\emptyset \tag{R6} \label{metabolite_consumption}\\
\text{mRNA} &\mathrel{\mathop{\rightarrow}^{\kdegr}} \emptyset \tag{R7} \label{mrna_degradation}\\
\text{Enzyme} &\mathrel{\mathop{\rightarrow}^{\mathrm{\kdil}}} \emptyset \tag{R8} \label{molecule_dilution1} \\
\text{Complex} & \mathrel{\mathop{\rightarrow}^{\mathrm{\kdil}}}\emptyset \tag{R9} \label{molecule_dilution2}
\end{align}

All reactions are assumed to follow mass action kinetics. Model simulations were computed with Gillespie's algorithm \cite{Gillespie1976} over long time horizons, in the order of hundreds of cell cycles for all simulations. Because of the complex multimodality observed, long simulations are needed to obtain accurate approximations of the stationary molecular distributions. The time courses shown in figures correspond to a small time window of the overall simulation. Unless mentioned in the figure captions, all parameter values were fixed to their nominal values shown in Table \ref{parameters_model1}. The parameters are selected in a physiologically realistic range respecting the scale separation of molecule numbers between mRNA ($\sim 1-5$ molecules), total enzyme abundance ($\sim 100$ molecules) and metabolites ($\sim 1000$ molecules).

To identify bimodality in \figref{bimodality_enzyme_product}, we detected the existence of one or two peaks in a distribution and defined it as bimodal if the height of the smaller peak is a least 10\% of the larger peak and the trough between peaks is at most 10\% of the height of the smaller peak.

\begin{table}[H]
\centering
\begin{tabular}[t]{lll}
   & value  & unit \\ \hline
\rowcolor{black!20}[\tabcolsep]$\substrate$ & 3000 &  molecule\\
$k_1$ & 1   &  $\text{s}^{-1} \text{molecule}^{-1}$\\
\rowcolor{black!20}[\tabcolsep]$k_{-1}$         & 1000   &  $\text{s}^{-1}$\\
					 $\kcat$ & 3.6  & $\text{s}^{-1} $\\
\rowcolor{black!20}[\tabcolsep] $\krev$ & 0.01 & $\text{s}^{-1} \text{molecule}^{-1}$ \\
$\kcons$         & 0.02    & $\text{s}^{-1} $
\end{tabular}
\,
\begin{tabular}[t]{lll}
    & value  & unit \\ 
\hline
\rowcolor{black!20}[\tabcolsep] $\ktransc$       & 0.0270 & $\text{s}^{-1} $\\
								$\ktransl$       & 0.2    & $\text{s}^{-1} $\\
 \rowcolor{black!20}[\tabcolsep] $\kon$ & 0.0225 &  $\text{s}^{-1} $\\
 $ \koff$ & 0.0075  & $\text{s}^{-1} $\\
\rowcolor{black!20}[\tabcolsep] $\kdegr$          & 0.2  &  $\text{s}^{-1} $ \\
$\delta$         & 0.00025  & $\text{s}^{-1} $ \\
\end{tabular}
\caption{\textbf{Nominal parameters for the stochastic model.} Parameters correspond to the simulations of Figure \ref{model_intro}A. We use realistic enzyme kinetic parameters \cite{Bar-Even2011} and fast promoter switching according to measured ranges \cite{So2011}. The dilution rate $\kdil$ constant corresponds to a doubling rate of approximately 46 minutes, typical in the \emph{Escherichia coli} bacterium. The distributions in Figure \ref{model_intro}C were obtained with perturbed promoter switching parameters $\kon=0.01 \text{ s}^{-1}$ and $\koff=\left\{0.03,\,0.01,\,0.0001\right\} \text{ s}^{-1}$. The parameter values for Figures \ref{bimodality_enzyme_product} and \ref{metabolite_multimodal} are shown in the respective captions.}
\label{parameters_model1}
\end{table}

\subsection*{Analytical expressions for the metabolite distribution}
To derive an analytic approximation for the probability to observe $\product$ metabolites in a cell, we first use the law of total probability as shown in~\eqref{PoissonMixtureModel}. 

\paragraph*{Distribution of the total enzyme ---}
Because free enzymes and complexes degrade at the same rate and $\etotal = \enzyme + \complex$ is conserved by the metabolic reaction, in the slow timescale the enzyme distribution $\mathbf{P}(\etotal)$ follows the standard solution\cite{Shahrezaei2008} of the three-stage model for gene expression:
\begin{align}
\label{eq:three_stage}
\mathbf{P}(\etotal) &=  \frac{\Gamma(\alpha_{+} + \etotal) \Gamma(\alpha_{-} + \etotal) \Gamma(\gamma)}{\Gamma(\etotal +1) \Gamma(\alpha_{+}) \Gamma(\alpha_{-}) \Gamma (\gamma + \etotal)}\times\notag\\ 
&\phantom{=}\left(\frac{b}{1+b}\right)^{\etotal} \left(1-\frac{b}{1+b}\right)^{\alpha_{+}}\times \notag \\
& \phantom{=}\textsubscript{2}F_{1} \left(\alpha_{+} + \etotal,\gamma - \alpha_{-}, \gamma +\etotal; \frac{b}{1+b} \right),
\end{align}
where $\Gamma$ is the Gamma function and $\textsubscript{2}F_{1}$ is the ordinary hypergeometric function. The parameters are $\gamma = (\kon+\koff)\slash\kdil$ and $\alpha_{\pm} = \left(a+\gamma \pm \sqrt{(a+\gamma)^2 - 4 a \koff\slash\kdil} \right)\slash2$, with $a = \ktransc\slash\kdil$ and $b = \ktransl\slash\kdegr$.

\paragraph*{Conditional distribution for the metabolite ---}
To compute the second term in~\eqref{PoissonMixtureModel}, we observe that enzyme expression occurs on a much longer timescale than enzyme kinetics, and thus metabolites can be considered to be in a quasi-equilibrium state of the catalytic reactions \eqref{substrate_binding}--\eqref{metabolite_catalysis} and metabolite consumption \eqref{metabolite_consumption}.
 
To explicitly compute the mixture components $\mathbf{P}(\product|\etotal)$, we assume that reversible binding between  substrate and enzyme in reaction \eqref{substrate_binding} is much faster than the catalytic step and metabolite consumption\cite{cao2005accelerated}. In this limit, the metabolite number evolves according to the effective reactions: 
\begin{align}
\label{eqn:slowreactions}
\emptyset \mathrel{\mathop{\rightleftarrows}^{\text{k}_{\text{cat}}^{\text{eff}}}_{\text{k}_{\text{rev}}^{\text{eff}}}}\text{Metabolite}\mathrel{\mathop{\rightarrow}^{\mathrm{\kcons}}} \emptyset,
\end{align}
where $\text{k}_{\text{cat}}^{\text{eff}}$ and $\text{k}_{\text{rev}}^{\text{eff}}$ are effective propensities averaged over the fast fluctuating variables $\complex$ and $\enzyme$:
\begin{align}
\begin{split}
	\text{k}_{\text{cat}}^{\text{eff}} &= \mathrm{\kcat} \mathbb{E}(\complex  |\etotal,\product),\\
	\text{k}_{\text{rev}}^{\text{eff}} &= \krev \mathbb{E}(\enzyme  |\etotal,\product),
	\end{split}\label{eff_propensities}
\end{align}
where $\mathbb{E}$ denotes the expectation operator. The derivation of the effective propensities in \eqref{eff_propensities} corresponds to a particular case of a more general methodology for timescale separation in stochastic chemical systems\cite{Goutsias2005, Thomas2014, Melykuti2014}. Note that since the total enzyme levels are conserved in the catalytic timescale, it follows that
\begin{align}
	\mathbb{E}(\enzyme  |\etotal, \product) +  \mathbb{E}(\complex|\etotal,\product) = \etotal.\label{etot_conserved}
\end{align}
To derive the conditional expectations in \eqref{eff_propensities}, we write the first-order moment equation for the free enzyme $\enzyme$, which according from equations \eqref{substrate_binding}--\eqref{metabolite_catalysis} reads
\begin{align}
&\frac{\mathrm{d} }{\mathrm{d}t} \mathbb{E}(\enzyme |\etotal, \product)= \notag\\
& -\kfor \substrate \mathbb{E}(\enzyme|\etotal,\product) + \kback \mathbb{E}(\complex|\etotal,\product) \notag\\
& + \kcat \mathbb{E}(\complex|\etotal,\product) - \krev \mathbb{E}(\enzyme|\etotal,\product) \product.\label{moment_equation}
\end{align}
Under the assumption that the reversible binding of substrate and enzyme is much faster than the other processes, the first two terms dominate the right hand side of equation \eqref{moment_equation} and determine the enzyme-complex quasi-equilibrium. Equating these two terms and using the conservation relation in \eqref{etot_conserved}, we obtain
\begin{align}\label{cond_means_supplement}
\begin{split}
\mathbb{E}(\enzyme  |\etotal,\product) &\approx \frac{ \kback }{(\kback+\kfor \substrate)} \etotal,\\
\mathbb{E}(\complex|\etotal,\product) &\approx \frac{\kfor \substrate}{\kback + \kfor\substrate} \etotal,
\end{split}
\end{align}
and thus both conditional expectations depend on $\etotal$ and are independent of the metabolite abundance. Therefore, the reactions in \eqref{eqn:slowreactions} correspond to a birth-death process with a zero-th order birth propensity $\text{k}_{\text{cat}}^{\text{eff}}$ and two linear death propensities. The mixture components $\mathbf{P}(\product|\etotal)$ are thus Poissonian with parameter $\lambda(\etotal)$ as shown in~\eqref{Metabolite_conditional_distribution}--\eqref{lambda}.

\subsection*{Comparison of PMM predictions and measured kinetic parameters}

The PMM depends on the effective parameters $\lambda_{\infty}$ and $K$, which are functions of five rate constants ($\kcat$, $\kfor$, $\kback$, $\kcons$ and $\krev$). Most of these parameters are not available, except $\kcat$ and $K_{\text{M}} = (\kcat + \kback)\slash\kfor$. From equation \eqref{effparameters} it follows
\begin{align}
\frac{\lambda_{\infty}}{K} = \frac{\kcat}{\kcons}\times \underbrace{\frac{\kfor \substrate}{\kfor \substrate + \kback}}_{\text{saturation }\epsilon},
\end{align}
which allows the computation of $\lambda_{\infty}\slash K$ for measured $\kcat$ values in different saturation conditions and consumption rate constants. The red line in \figref{figure_parameter_value}B represents the $\lambda_{\infty}\slash K$ ratio for a saturated enzyme ($\epsilon = 1$), fast consumption ($\kcons = 100\times\delta$) and the median $\kcat \approx 16.5 \text{ s}^{-1}$ in \textit{E. coli} \cite{Bar-Even2011}. The top grey line is the case without consumption, \ie metabolites are diluted by cell growth ($\kcons=\delta$). Lower saturation ($\epsilon = 0.2$) moves the red line down the vertical axis (bottom grey line). The enzyme distribution in \figref{figure_parameter_value}B was produced with promoter switching parameters $\left\{\kon,\,\koff\right\} = \left\{1.56,\,3\right\}\times10^{-4}\text{ s}^{-1}$, $\ktransc=0.025\text{ s}^{-1}$, and $\ktransl = 0.2\text{ s}^{-1}$. The boundaries between unimodal, bimodal and multimodal distributions were computed as follows. Unimodal distributions are those with a single maximum. Bimodal distributions were detected as in Figure \ref{bimodality_enzyme_product}. Multimodal distributions are those with at least one additional peak higher than a threshold of $1\times 10^{-4}$ and the trough between neighbouring peaks at most 90 \% of its height.

\section*{Acknowledgements}
This work was funded by the Human Frontier Science Program through a Young Investigator Grant awarded to D.O.~(RGY0076-2015), The Royal Commission for the Exhibition of 1851 through a Fellowship to P.T., and the EPSRC Centre for Mathematics of Precision Healthcare (EP/N014529/1).

\section*{Author contributions}
M.T., P.T., M.B and D.O. conceived the study. M.T. carried out the theoretical derivations, simulations and analysed the data. M.B. and D.O. supervised the work. M.T., P.T., M.B. and D.O. wrote the paper.

\section*{Competing interests}
The authors declare no competing interests.

\section*{Code availability}
The code for producing model simulations is available from the authors upon request.

\section*{Data availability}
All data is available from the authors upon request.

\section*{References}

\end{document}

%% file: header.tex
\usepackage{amsmath}
\usepackage{amsfonts}
\usepackage{amssymb}
\usepackage{amsxtra}
\usepackage{floatflt}
\usepackage{comment}
\usepackage[pdftex]{graphicx}
\usepackage{enumerate}
\usepackage{float}
\usepackage[normalem]{ulem}
\usepackage{color}
\usepackage{mathtools}
\graphicspath{{./figs/}}
\usepackage{marvosym}
\usepackage{times}
\usepackage{bm}
\usepackage{colortbl} 
\usepackage{multirow}
\usepackage{graphicx}
\usepackage{nameref}
\usepackage[usenames,dvipsnames,svgnames,table]{xcolor}
\usepackage{float}
\usepackage[usestackEOL]{stackengine}

\usepackage{tabularx}
\newcolumntype{L}[1]{>{\raggedright\arraybackslash}p{#1}} 
\newcolumntype{C}[1]{>{\centering\arraybackslash}p{#1}} 
\newcolumntype{R}[1]{>{\raggedleft\arraybackslash}p{#1}} 

\newcommand{\eg}{e.g.~}
\newcommand{\ie}{i.e.~}

\newcommand{\figref}[1]{Figure \ref{#1}}

\newcommand{\kon}[1]{\kappa^{\mathrm{1}}_{#1}}
\newcommand{\koff}[1]{\kappa^{\mathrm{0}}_{#1}}

\newcommand{\kcat}[1]{k_{\mathrm{cat}\,#1}}
\newcommand{\Km}[1]{K_{\mathrm{M}\, #1}}

\newcommand{\enzyme}{n_{\text{e}}}
\newcommand{\complex}{n_{\text{c}}}
\newcommand{\product}{n_{\text{p}}}
\newcommand{\substrate}{n_{\text{s}}}
\newcommand{\etotal}{n_{\text{etot}}}

\newcommand{\don}{D_{\text{on}}}
\newcommand{\doff}{D_{\text{off}}}

\newcommand{\kfor}{k_{1}}
\newcommand{\kback}{k_{-1}}
\renewcommand{\kcat}{k_{\text{cat}}}
\newcommand{\krev}{k_{\text{rev}}}
\newcommand{\kdil}{\delta}
\newcommand{\kcons}{k_{\text{c}}}

\newcommand{\ktransc}{k_{\text{tx}}}
\newcommand{\ktransl}{k_{\text{tl}}}
\newcommand{\kdegr}{k_{\text{deg}}}

\renewcommand{\kon}{k_{\text{on}}}
\renewcommand{\koff}{k_{\text{off}}}